\documentclass[amsmath,amssymb,nofootinbib,showpacs,twocolumn,nofootinbib]{revtex4-1}
\usepackage{dcolumn}
\usepackage{bm}
\usepackage{graphicx}
\usepackage{epstopdf}
\usepackage{epsfig}
\usepackage{color}
\usepackage{wasysym}
\usepackage{natbib}
\usepackage{twoopt}
\usepackage{dashrule}
\usepackage[breaklinks=true,colorlinks=true,linkcolor=blue,citecolor=blue,urlcolor=blue,pdfauthor={Tomassetti},pdftitle={Wanted}]{hyperref}
\def\MyTitle#1{{\textit{#1}}}
 
\definecolor{ColorTitle}{cmyk}{0,.88,.77,.40}

\newcommand{\ApJ}{ApJ}
\newcommand{\AeA}{A\&A}
\newcommand{\PRL}{PRL}
\newcommand{\PRD}{PRD}
\newcommand{\JCAP}{JCAP}
\newcommand{\PLB}{PLB}

\newcommand{\AMS}{\textsf{AMS}}
\newcommand{\etal}{et al.}
\newcommand{\eg}{\textit{e.g.}} 
\newcommand{\ie}{\textit{i.e.}} 

\newcommand{\R}{\ensuremath{\mathcal{R}}}
\newcommand{\p}{\textsf{p}}
\newcommand{\He}{\textsf{He}}
\newcommand{\Li}{\textsf{Li}}
\newcommand{\Be}{\textsf{Be}}
\newcommand{\B}{\textsf{B}}
\newcommand{\C}{\textsf{C}}
\newcommand{\N}{\textsf{N}}
\newcommand{\Oxy}{\textsf{O}}
\newcommand{\Fe}{\textsf{Fe}}
\newcommand{\BC}{\textsf{B}/\textsf{C}}
\newcommand{\LiO}{\textsf{Li/O}}
\newcommand{\TiFe}{\textsf{Ti/Fe}}

\newcommand{\epfrac}{e\ensuremath{^{+}}/(e\ensuremath{^{-}}\,+\,e\ensuremath{^{+}})} 
\newcommand{\epm}{\ensuremath{e^{\pm}}}
\newcommand{\pbarp}{\textsf{\ensuremath{\bar{p}/p}}}

\newcommand{\Dragon}{\texttt{DRAGON}}

\newcommand{\CRDB}{\texttt{CRDB}}
\newcommand{\Paper}{paper}

\begin{document}
\title{Cosmic-ray protons, nuclei, electrons, and antiparticles\\ under a two-halo scenario of diffusive propagation}
\author{Nicola Tomassetti}
\address{LPSC, Universit\'e Grenoble-Alpes, CNRS/IN2P3, F-38026 Grenoble, France; email: nicola.tomassetti@lpsc.in2p3.fr}
\date{September 2015} 
\begin{abstract}
We report calculations of cosmic-ray proton, nuclei, antiproton, electron and positron energy spectra within a ``two-halo model'' 
of diffusive transport. The two halos represent a simple, physically consistent generalization of the standard diffusion models, 
which assume a unique type of diffusion for cosmic rays in the whole Galactic halo.
We believe instead that cosmic rays may experience a smaller energy dependence of diffusion when they are in proximity of the Galactic disk.
Our scenario is supported by recent observations of cosmic ray protons, nuclei, anisotropy, and gamma-rays. 
We predict remarkably 
hard antiparticle spectra at high energy. In particular, at $E\gtrsim$\,10 GeV, the antiproton/proton ratio is expected to
flatten, while the positron fraction is found to increase with energy. We discuss the implications for cosmic-ray physics 
and dark matter searches via antimatter.
\end{abstract}
\maketitle

\MyTitle{Introduction} --- 
%
The observed spectrum of Galactic cosmic rays (CRs) is believed to arise from a combination of two basic plasma 
astrophysics phenomena: diffusive shock acceleration occurring in supernova remnants,
and diffusive transport off magnetic turbulence \citep{Grenier2015,Bell2014,Blasi2013,Strong2007}. 
The paradigm of CR transport as a particle diffusion process has been successful for decades in
explaining the observed high-degree of isotropy in CR arrival directions, and the abundance of 
secondary \Li-\Be-\B{} elements arising from CR collisions with the interstellar matter.
The so-called \emph{standard models} based on this understanding employ several simplifying assumptions such as 
homogeneity, isotropy, stationarity, and linearity. 
The traditional description is that CRs are confined in a cylindrical halo 
that encompasses the Galactic plane, after being accelerated by supernova remnants to power-law spectra
in rigidity $\R$ (momentum/charge ratio), $Q\propto \R^{-\nu}$. 
The diffusion coefficient is assumed to have a rigidity dependence $K\propto \R^{\delta}$ in the whole halo.
The combination of acceleration and transport leads to power-law equilibrium spectra 
$\phi_{p} \sim Q/K \propto \R^{-\delta-\nu}$ for primary components (\eg, protons, \He, \C, \Oxy, \Fe),
while secondary to primary ratios (\eg, \BC, \LiO, \TiFe) scale as $\phi_{s}/\phi_{p}\sim \R^{-\delta}$. 
The data on primary nuclei and on the  \BC{} ratio constrain the parameters $\delta\approx$\,0.3--0.7 and $\nu\approx$\,2--2.4. 

New generation experiments, and in particular the Alpha Magnetic Spectrometer (\AMS), are now probing the fine structure of CR phenomenology. 
The rapid rise in the positron fraction \epfrac{} at $E\sim$10--300 GeV of energy \citep{Accardo2014,Adriani2009}, in contrast
to the standard expectations $\sim E^{-\delta}$, may suggest the existence of extra sources of 
high-energy $e^{\pm}$, including dark matter particle annihilation/decay or known astrophysical sources,
but these scenarios demand a reliable understanding the secondary \epm{} production and their propagation
\citep{Choi2014,Li2015,MertschSarkar2014,TomassettiDonato2015,Blum2013,Serpico2012}.
The remarkable spectral hardening observed in CR protons and nuclei at $\R\sim$\,300\,GV \citep{Adriani2011,Yoon2011,Panov2009,Aguilar2015}
may be a signature of new astrophysical phenomena, occurring either in acceleration or in propagation, that are unaccounted by 
standard models \citep{Ptuskin2013,Tomassetti2012,Blasi2012,Vladimirov2012,ThoudamHorandel2014}.
Revisiting propagation, as distinct from acceleration, has also interesting implications for the so-called
``anisotropy problem'' \citep{Pohl2013}. 
The anisotropy observations suggest a rather small rigidity dependence for the diffusion at multi-TeV energies ($\delta\approx$\,0.15),  
which is in contrast with standard model extrapolations based on the \BC{} ratio ($\delta\approx$\,0.3--0.7).
Yet a change of diffusion may also be hinted by a high-energy flattening of secondary to primary ratios \citep{Pohl2013,Evoli2014,Blasi2012}.
Very recently, new \pbarp{} data presented by \AMS{} have generated widespread interest \citep{AMSDays2015,Chen2015,Kohri2015,Hamaguchia2015,Kachelriess2015},
and some authors have pointed out that the interpretation of these data in terms of secondary production demands a
slower diffusion of cosmic rays at high energy  \citep{Giesen2015,Kappl2015,Cowsik2013}. 
We also stress that understanding diffusion is crucial, in dark matter searches, for modeling \emph{both} background and signal of CR antimatter.

In this \Paper, we report calculations of CR proton, nuclei, antiproton, electron and positron 
energy spectra within a numerical implementation of a \emph{two-halo} scenario of diffusive transport. 
The idea of having two halos with different diffusion properties represents the simplest but physically consistent 
generalization of the standard diffusion models. Standard models assume a unique diffusion regime in the whole 
propagation region, \ie, a factorized rigidity dependence for the function $K$.
The increased turbulence in the disk, compared with that at higher latitudes, leads 
us to believe that there is a spatial change in the CR diffusion properties.
This idea was proposed in connection with the detection of a break in the proton and \He{} spectra \citep{Tomassetti2012},
but the model was unable to describe the sharp structures reported by the PAMELA experiment. 
The \AMS{} Collaboration has now measured the detailed variations of the proton and \He{} fluxes at 
GeV--TeV energies, showing a significantly smoother spectral hardening.
In particular, the proton spectral index is found to progressively increase at $E\gtrsim$\,100\,GeV \citep{Aguilar2015}.
As we will show, a two-halo scenario of CR diffusion is supported by these data, 
and it allows us to understand several puzzling properties of the CR spectrum.
Throughout this \Paper, we discuss important consequences for the physical 
observables that are being investigated by \AMS, including implications for dark matter searches via antimatter.

\MyTitle{Numerical Implementation} --- 
%
The CR diffusion is described by a 2D transport equation with boundary conditions in a cylindrical region for all CR species:
\begin{equation}\label{Eq::DiffusionTransport}
  \partial_{t} {\psi} = Q + \vec{\nabla}\cdot (D\vec{\nabla}{\psi}) - {{\psi}}{\Gamma} + \partial_{E} (\dot{E} {\psi})  \,,
\end{equation}
where $\psi=\psi(E,r,z)$ is the particle number density as a function of energy and space coordinates,
$\Gamma= \beta c n \sigma$ is the destruction rate for collisions off gas nuclei, with density $n$,
at velocity $\beta c$ and cross section $\sigma$. The source term $Q$ is split into a primary term, $Q_{\rm pri}$, 
and a secondary production term $Q_{\rm sec}= \sum_{\rm j} \Gamma_{j}^{\rm sp} \psi_{\rm j}$, from spallation of 
heavier $j$--type nuclei with rate $\Gamma_{j}^{\rm sp}$. 
The term $\dot{E}=\--\frac{dE}{dt}$ describes ionization and Coulomb losses, as well as radiative cooling of CR leptons.
To numerically solve Eq.\,\ref{Eq::DiffusionTransport} in steady state conditions ($\partial_{t}\psi$\,=0), 
we employ the \Dragon{} package \citep{Dragon}. 
We introduced a modification of the finite-differencing scheme and of the spatial grid of the code in order to allow for a
diffusion coefficient of the form
$K(\R,z)= K_{0}\left( \R/\R_{0}\right)^{\delta(z)}$. 
Note that our function $K$ is \emph{non-separable} in rigidity and space coordinates.
The full propagation region has half-height $L=$\,4\,kpc and radius $r_{\rm max}=20$\,kpc. We split the cylinder into two $z$-symmetric propagation zones. 
We call \emph{inner halo} the region which surrounds the disk for a few hundred pc 
($|z|<\l_{i}\cong$\,500\,pc), where the turbulence is presumably injected by SN bubbles,
while the \emph{outer halo} represents a wider region ($l_{i}<|z|<L$) 
possibly characterized by a CR-driven turbulence spectrum \citep{ErlykinWolfendale2002}.
The spatial dependence of $\delta(z)$ is expressed by $\delta=\delta_{i}\cong$\,0.15 in the inner halo and $\delta=\delta_{o}\cong$\,0.75 in the outer halo, 
where $K_{0}\cong$10$^{28}$\,cm$^{2}$s, and $\R_{0}\equiv$\,0.25\,GV set the diffusion in the two zones. 
The source spectra are of the type $Q^{\rm pri}\propto \R^{-\nu}$, with $\nu$=2.32 for nuclei, while proton spectra are steeper by 0.08 \citep{Malkov2011}.
For illustrative purposes, we also set up a \emph{standard model} of 
Iroshnikov-Kraichnan type diffusion,
with $\delta\equiv$\,0.5 everywhere and source spectra with $\nu=$\,2.3, to match the data at $E\sim$\,1--200\,GeV/nucleon.
The local interstellar flux is $\phi=\frac{\beta c}{4\pi}\psi_{\odot}$, computed at the position $z_{\odot}$=0 and $r_{\odot}=$8.3\,kpc.
In the following, the particle fluxes at Earth from the two models are presented. 
The solar modulation is described under the \emph{force-field} approximation \citep{Gleeson1968}, using modulation parameter values $\Phi\cong$\,550\,MV for protons and leptons (relevant for the \AMS{} data) and $\Phi\cong$\,350\,MV for nuclei (relevant for the PAMELA data) \citep{Maurin2014,Maurin2015}.

\begin{figure}[!t]
\includegraphics[width=0.44\textwidth]{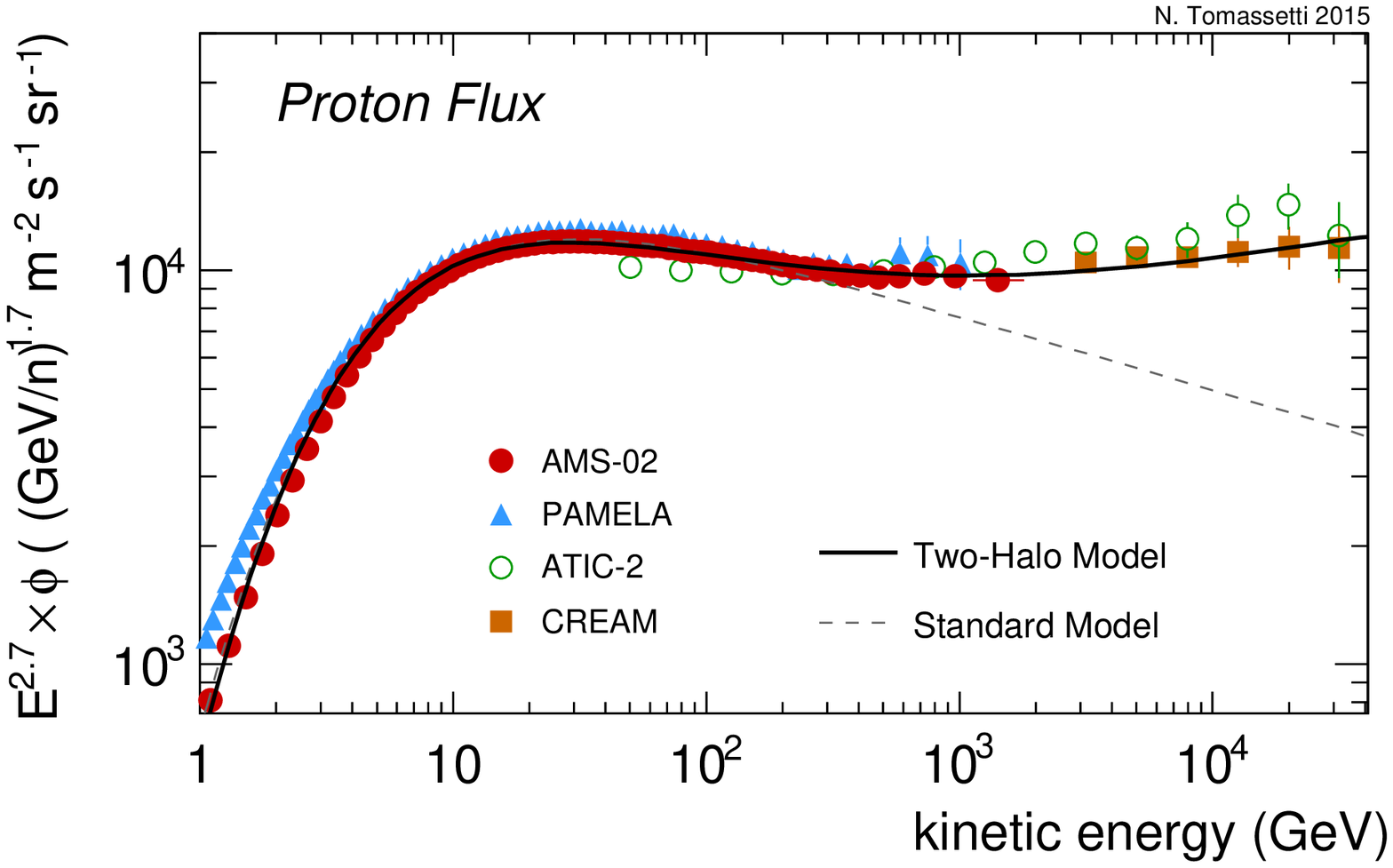}
\includegraphics[width=0.44\textwidth]{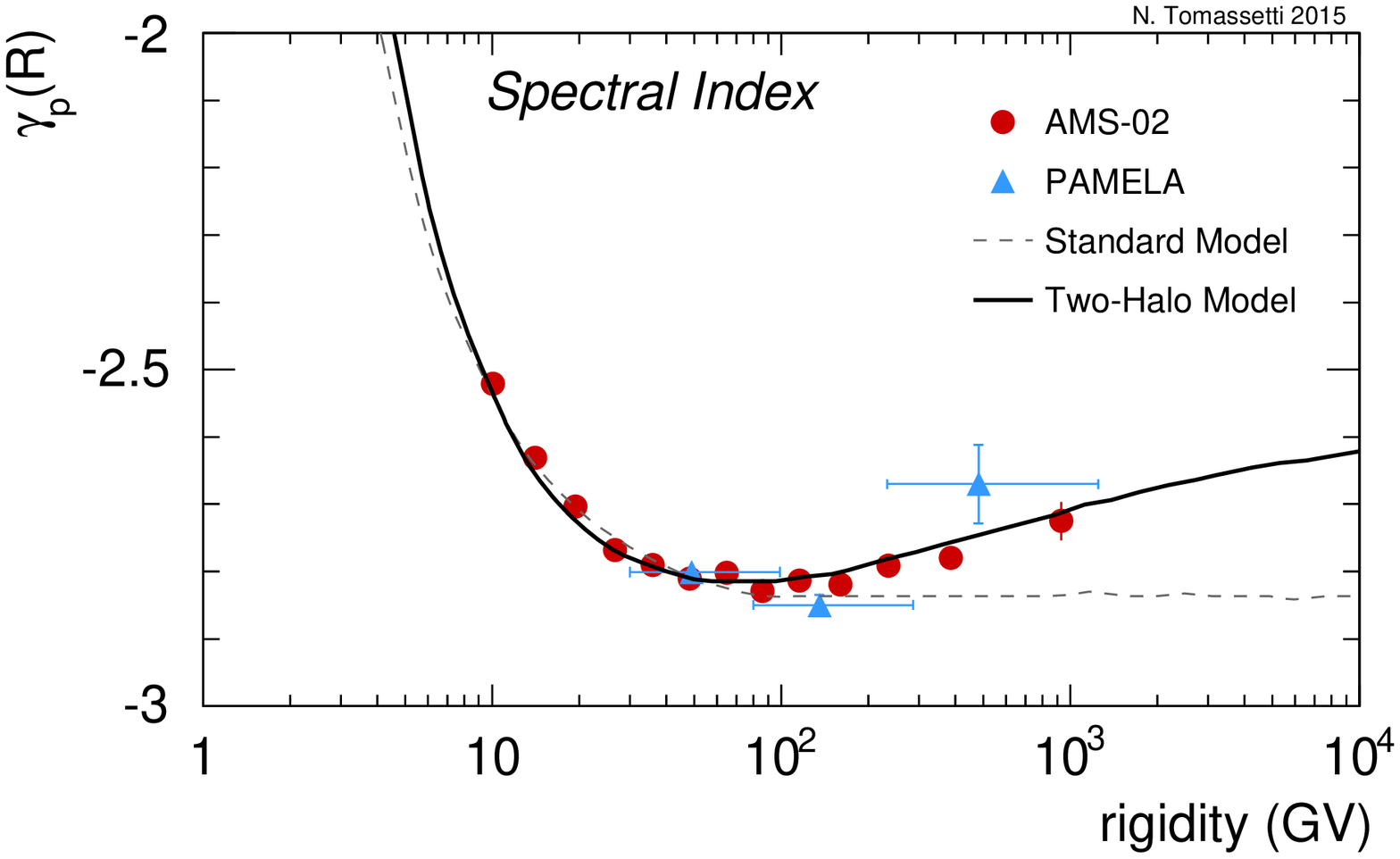}
\caption{ 
  Top: proton energy spectrum multiplied by $E^{2.7}$. Bottom: rigidity dependence of the proton spectral index. 
  The model calculations are shown in comparison with the data \citep{Aguilar2015,Adriani2011,Yoon2011,Panov2009}.
  The modulation potential is $\Phi=550$\,MV.
}
\label{Fig::ccProtonSpectrum}
\end{figure}

\MyTitle{Protons} --- 
%
The calculated proton spectrum is shown in Fig.\,\ref{Fig::ccProtonSpectrum} in comparison with the new \AMS{} data.
At $E\gtrsim$\,10\,GeV, CR protons are not very sensitive to energy losses or destruction.
Our numerical results nicely confirm the trends obtained in our analytical derivations \citep{Tomassetti2012}.
Neglecting interactions, one can see that the naive expectation, $\phi_{p}\sim\,Q^{\rm pri}/\Omega$ with
$\Omega=K/L$, has to be modified using $\Omega=\left( l_{i}/k_{i} + l_{o}/k_{o} \right)$, 
which depends on the diffusion coefficients $k_{i/o}\propto \R^{\delta_{i/o}}$ and on the half-sizes of the two halos $l_{i/o}$.
The CR propagation in both regions, with systematically slower diffusion in the inner halo,
is at the origin of the spectral hardening.
The high-energy propagation becomes progressively dominated by the inner halo.
In the bottom panel, we plot the spectral index $\gamma(\R)=d[\log(\phi_{p})]/d[log(\R)]$.
Our model describes very well the smooth evolution of the spectrum 
measured by \AMS{} at $\R\sim$\,10-1000\,GV 
We also note that, in this rigidity region, the spectrum is never power-law. 

\begin{figure}[!t]
\includegraphics[width=0.44\textwidth]{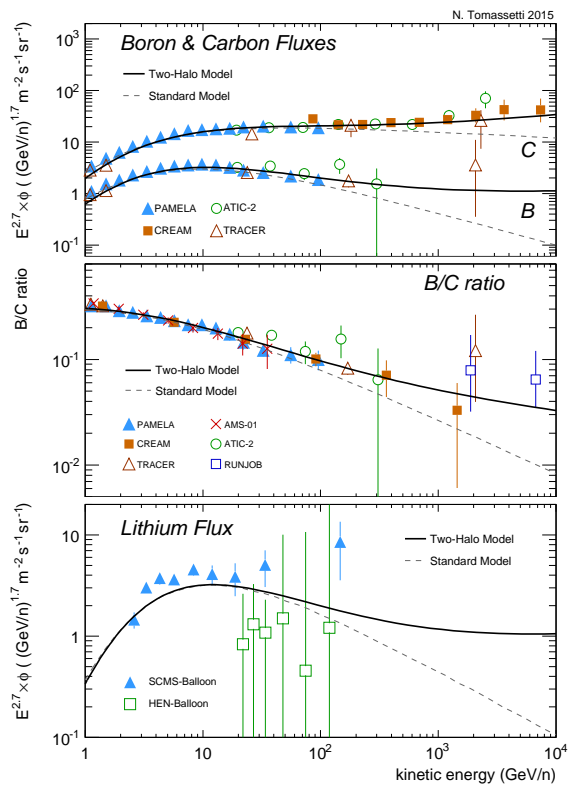}
\caption{ 
  Top to bottom: Energy spectra of \B{} and \C, \BC{} ratio, and \Li{} 
  spectrum. The spectra are multiplied by $E^{2.7}$.
  The model calculations are shown in comparison with the data 
  \citep{Adriani2014,Ahn2010,Obermeier2012,Panov2009,Aguilar2010,Derbina2005,Orth1978,Juliusson1974}.
  The modulation potential is $\Phi=350$\,MV.
}\label{Fig::ccNucleiSpectra}
\end{figure}

\MyTitle{Nuclei} --- 
%
The transport of primary nuclei (\eg, \He, \C, \Oxy{} or \Fe) is similar to that of protons,
hence we expect a spectral hardening at the same rigidity as for proton.
A minor difference comes from destructive interactions, increasing with increasing masses, 
that slightly harden the spectra below $\sim$\,100\,GeV/nucleon. 
The carbon spectrum is shown in Fig.\,\ref{Fig::ccNucleiSpectra}. 
At $E\gg$\,GeV/n, its approximate behavior is $\phi\sim {Q}/\left( \Omega + h\Gamma \right)$,
where  $h\Gamma$ is the interaction term for a surface gas density $hn \cong$\,200\,pc\,$\times$\,1\,cm$^{-3}$,
acting in competition with the diffusive term $\Omega$.
%
Secondary nuclei, like boron in the figure, are expected to have steeper spectra and to experience a stronger 
upturn in comparison to their progenitors \C-\N-\Oxy. 
As a consequence the \BC{} ratio is found to progressively flatten, as shown in the figure, 
to an asymptotic multi-TeV dependence $\sim E^{-\delta_{i}}$. At lower energies, the \BC{} ratio from our model recovers the standard model behavior. 
A hint of flattening in the \BC{} data is noted in several studies  \citep{Pohl2013,Evoli2014,Blasi2012,TomassettiDonato2012,MertschSarkar2014}.
The lithium flux, shown in Fig\,\ref{Fig::ccNucleiSpectra}, is particularly sensitive to CR propagation.
Its production depends not only on \C-\N-\Oxy{} collisions, but also on \emph{tertiary} interactions \B$\rightarrow$\Li{} and \Be$\rightarrow$\Li.
In spite of the scarcity of present data, the \Li{} flux is one of the best observables
for \AMS{} in terms of detection performance, and it is currently being measured to a $\sim$\,1\% precision up to TeV/nucleon 
energies \citep{AMSDays2015}. These data will be very valuable for testing our scenario. 
%
 
\begin{figure}[!t]
\includegraphics[width=0.44\textwidth]{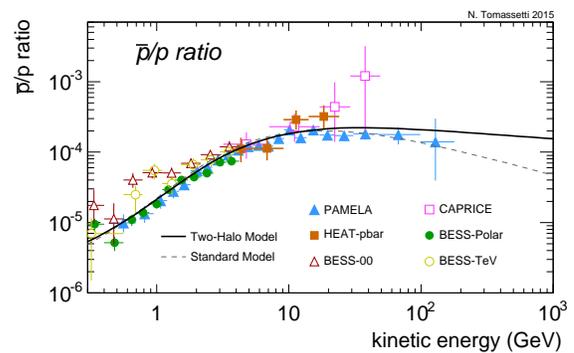}
\caption{ 
  The \pbarp{} ratio as function of kinetic energy. The model calculations
  are shown in comparison with the data \citep{Adriani2010,Beach2001,Boezio2001,Abe2008,Asaoka2002,Haino2005}. 
  The modulation potential is $\Phi=350$\,MV.
}
\label{Fig::ccPbarPRatio}
\end{figure}

\MyTitle{Antiprotons} --- 
%
In contrast to \Li-\Be-\B{} nuclei, that are ejected with the same kinetic energy per nucleon 
as their progenitors, the antiprotons emitted in \p-\p{} or \p-\He{} collisions have broad energy 
distributions and large inelasticity factors.
This type of kinematics reshapes significantly the antiproton production spectrum $Q_{\bar{p}}^{\rm sec}$. 
Their subsequent propagation is similar to that of protons, except for the presence of
the so-called \emph{tertiary} $\bar{p}$--$p$ processes.
In Fig.\,\ref{Fig::ccPbarPRatio} the \pbarp{} ratio is shown for the two considered models. 
In the standard model, it decreases smoothly above $\sim$\,10\,GeV.
In our two-halo model, the ratio has a substantial flattening at $E\sim$\,10\,GeV -- a few 100\,GeV. 
With the present data, it appears that no room is left for exotic antiproton components. 
Recent \AMS{} data have triggered exciting speculations about a possible 
dark matter induced antiproton excess  \citep{Chen2015,Kohri2015,Hamaguchia2015}.
A flattening of the \pbarp{} ratio is an unavoidable feature of our scenario and 
it should be considered when setting upper limits to the search for new-physics signals.

\MyTitle{Anisotropy} ---  
%
The CR anisotropy has received renewed attention in recent years \citep{Pohl2013,Savchenko2015,Kumar2014}. 
From observations, the dipole anisotropy in the TeV band remains nearly constant to $\eta\sim$\,10$^{-3}$. 
Standard model predictions exceed the observational limits by one order of magnitude. 
To match the data, a diffusion coefficient with $\delta\approx$\,0.15 is required. This is indeed the case for
the model considered here, which gives $\eta\lesssim$\,2$\times$10$^{-3}$ at 1--100\,TeV. 
These estimates depend on the standard formula for anisotropy, 
$\eta\equiv\frac{3K}{c} \frac{\nabla\psi}{\psi}$,
within the diffusion approximation.
We also stress that accounting for the discreteness of the CR sources may induce large deviations from these estimates.

\MyTitle{Connections with $\gamma$-ray physics} --- 
%
Although we have focused on \emph{local} observables, our model predicts 
also a CR spectral steepening with increasing distance from the disk.
Moreover, one may also argue for a smooth \emph{radial} dependence for the parameter $\delta$. 
Some theoretical arguments are found in \citep{ErlykinWolfendale2002,Evoli2014}.
Observationally, a spectral steepening toward high latitudes or longitudes has been found
in $\gamma$--rays by the \emph{Fermi}-LAT observatory \citep{Ackermann2012}.
As argued in \citep{ErlykinWolfendale2013} and very recently 
shown in \citep{Gaggero2015}, the  \emph{Fermi}-LAT
data are suggestive of a spatial change of CR diffusion. 
These works are nicely complementary to the present paper.

\begin{figure}[!t]
\includegraphics[width=0.44\textwidth]{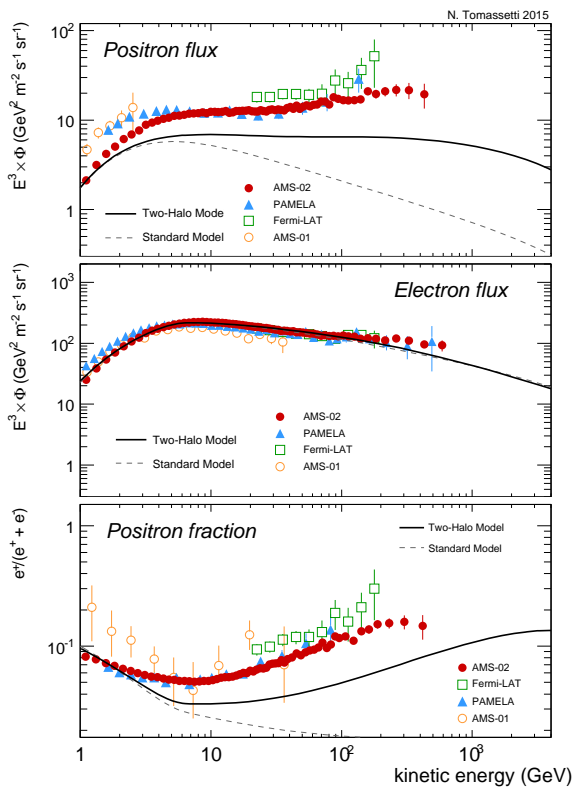}
\caption{ 
  Energy spectra of positrons and electrons
  multiplied by $E^{3}$, and positron fraction \epfrac.
  The model calculations are shown in comparison with the data 
  \citep{Accardo2014,Adriani2009,Aguilar2014,Adriani2013,Aguilar2007,Abdo2012}. 
}
\label{Fig::ccLepton}
\end{figure}

\MyTitle{Leptons} ---  
%
In contrast to hadrons, light CR leptons are subjected to synchrotron radiation and inverse Compton losses with a characteristic 
time-scale $\tau(E) \propto E^{-1}$. These processes limit the range that they can travel to 
distance scales $\lambda_{i/o} \sim \sqrt{\tau k_{i/o}}$, in the two propagation regions. Above a few GeV one has $\lambda_{o}\lesssim l_{o}$ 
in the outer halo so that, in contrast to hadrons, CR leptons detected at Earth have spent a larger 
fraction of time in the inner halo. At higher energies the distance scale
decreases further. The resulting effect is that the $e^{\pm}$ spectra above a few 
tens of GeV are essentially determined by the propagation in the inner halo. 
Radiative losses also steepen the $e^{\pm}$ spectrum with a stronger effect than diffusion.
The lepton fluxes at $E\gtrsim$\,10\,GeV are of the type $\phi_{e^{\pm}}\sim (\tau/K)^{1/2}Q$ and 
their spectral index scales as $\gamma_{e}\approx -\nu_{e} - \frac{1}{2}(\delta+1)$ where, for the 
two-halo scenario, we have $K\cong\,k_{i}$ and $\delta\cong\delta_{i}$. 
In summary, the leptonic spectra under the two-halo model are expected to be 
harder, due to propagation, by a factor $\frac{1}{2}(\delta-\delta_{i}) \approx$\,0.15--0.2. 
In addition, secondary $e^{\pm}$ have a further hardening factor coming from the spectra of their parent nuclei \p-\He.
The results of our numerical calculations are shown in Fig.\,\ref{Fig::ccLepton}. 
The force-field approximation is inadequate for CR leptons \citep{Maccione2013}, hence we focus on $E\gg$\,GeV. 
It can be seen that our model gives an enhanced flux of secondary positrons at high energy.
For primary electrons we expect a similar effect, but their flux depends on the adopted injection index $\nu_{e}$, which is not known a priori. 
Using the quantity $\Delta\Phi_{e^{\pm}}\equiv \phi_{e^{-}}-\phi_{e^{+}}$ to minimize the secondary and extra components \citep{Jin2015}, 
we derived $\nu_{e}\cong$\,2.6 for the standard model and $\nu_{e}\cong$\,2.77 for the two-halo model.
Hence the propagated electron fluxes for the two models are similar.
It is interesting to note that our model, with the adopted parameters, matches the conditions
for having a positron fraction that \emph{increases} with energy. 

As for the case of antiprotons, a high-energy enhancement of the positron flux is a
characteristic feature of the two-halo scenario of CR diffusion.
The high-energy positron flux is significantly larger than that arising from standard models. 
For dark matter models with TeV scale masses, 
the increased level of \emph{background}, together with a much slower 
propagation of the \emph{signal}, will changes the intensity and the spectral shape of the expected $e^{\pm}$ yield from exotic sources.
At this point, one may be tempted to explain the \epm{} data entirely with
a purely diffusive picture.
However, we found that even the extreme choice $\delta_{i}\cong$\,0 is unsatisfactory for CR leptons. 
Besides, 
this would require exceedingly steep injection spectra and a worse agreement with the \BC{} ratio data.
Our choice of $\delta_{i}\cong 0.15$ represents at present the best agreement between \BC{} 
data and anisotropy requirements. Nonetheless, the current \BC{} data are not very constraining.
Light-nuclei data from \AMS{} will allow for much more stringent predictions.

\MyTitle{Conclusions} --- 
%
Several observed properties of the CR spectrum can be understood in terms of spatial changes of CR diffusion in the Galaxy, 
in contrast to standard propagation models that assume a unique diffusion regime everywhere. 
In our scenario, a departure from the standard power-law form is predicted for all elemental fluxes. 
We found an excellent agreement with the new \AMS{} proton data at $\sim$\,1\,GeV -- 2\,TeV of energy.
Our model can reconcile the steeply falling low-energy shape of the \BC{} ratio with the high-energy requirements of the anisotropy data. 
In this we have devised perhaps one of the few viable solutions for the \emph{anisotropy problem}. 
Our model shares some consequences with the model proposed in \citep{Blasi2012}.
The $\gamma$-ray diffuse emission may be useful to discriminate the two scenarios. 
Our predictions have also implications for the Galactic synchrotron emission, 
which traces the electron spectrum at different latitudes \citep{OrlandoStrong2013}.
Interesting results are found for the antiparticle spectra, especially in view of recent excitement about possible evidence for dark matter. 
Our model predicts remarkably hard spectra for secondary antiproton and positron at high energy.
In particular, the \pbarp{} ratio is found to flatten above $\sim$\,10\,GeV, while the positron fraction is found to increase with energy. 
These results have a direct impact on upper limits on new physics signals using the \pbarp{} ratio, 
or on dark matter models for the positron fraction. The antimatter spectra are strictly related to secondary nuclei.
The main limitation is represented by the large uncertainties of the pre-\AMS{} nuclear data.
A more robust assessment will be done upon the forthcoming release of the \AMS{} data on \He{} \Li, \C, \BC{} and \pbarp.
In particular, the \Li{} flux will be a powerful tool at the level of precision expected by \AMS.
If new data confirms our scenario, then the current \emph{standard models} of CR diffusion will have to be revised to account for it. \\[0.3cm]
{\footnotesize%
  I thank Jin Chao and Carmelo Evoli for discussions and support with \Dragon, 
  Sherry Chou for reading this manuscript,
  and the \Dragon{} team for sharing their code with the community. 
  Several data are extracted from the \CRDB{} database \citep{Maurin2014}. 
  This work is supported by the ANR LabEx grant \textsf{ENIGMASS} at CNRS/IN2P3.
}



\end{document}